\documentclass{article}
\usepackage{amsfonts}
\usepackage{amssymb}
\usepackage{amsmath}
\usepackage{amsthm}
\usepackage[english]{babel}
\begin{document}

\title{\bf The Variational principle, \\ Conformal and Disformal transformations, \\ and the degrees of freedom}

\author{{\bf Alexey Golovnev}\\
{\small {\it Centre for Theoretical Physics, The British University in Egypt,}}\\
{\small \it El Sherouk City, Cairo 11837, Egypt}\\
{\small agolovnev@yandex.ru}}
\date{}

\maketitle

\begin{abstract}

Conformal and disformal transformations are now being very intensively studied in the context of various modified gravity theories. In particular, some special classes of them can be used for constructing Mimetic Dark Matter models. Recently, it has been shown that many more transformations of this type, if not virtually all of them when the coefficients depend on the scalar kinetic term, can produce new solutions with mimetic properties. The aim of this paper is to explain how it works at the level of the variational principle, and to express some worries about viability of these models.

\end{abstract}

\section{Introduction}

We live in the era of active modified gravity research, with various motivations. Even though it is quite hard to claim any undoubtful success in that, there are many very interesting theories on the market. In particular, some time ago, a model of mimetic gravity was proposed \cite{mimetic}. It consists of the standard Einstein-Hilbert action of General Relativity but with the physical metric represented as a composite field $g_{\mu\nu}={\tilde g}_{\mu\nu}{\tilde g}^{\alpha\beta}(\partial_{\alpha}\phi)(\partial_{\beta}\phi)$ where the auxiliary metric $\tilde g$ and a scalar field $\phi$ are the new fundamental dynamical variables. 

Of course, what has happened there is simply that extra variables were introduced, meaning that some combinations of them will be unphysical, however the variational principle still works in terms of what can be taken as the stationary action condition with respect to arbitrary perturbations of the physical metric $g$ represented in terms of a bigger set of quantities. Nevertheless, the equations of motion turn out to allow for more general solutions than in pure General Relativity, in the shape of an irrotational pressureless dust,  dubbed Mimetic Dark Matter \cite{mimetic}, added to the right hand side of the Einstein equations. Why is that?

In this particular case, the reason is actually obvious \cite{my}. By its very definition, the auxiliary metric enjoys the local conformal symmetry ${\tilde g}_{\mu\nu}\longrightarrow \alpha(x)\cdot {\tilde g}_{\mu\nu}$. In other words, its conformal mode is precisely the piece of redundant information due to overproduction of variables for describing the physical metric geometry. What it means for the physical metric $g_{\mu\nu}$ in the action is that its conformal factor and variations thereof cannot be represented in terms of the auxiliary metric ${\tilde g}_{\mu\nu}$. On the contrary, it's given solely in terms of gradients of the scalar field $\phi$. The gradients being extra derivatives introduced into the action make the equations of motion allow for more solutions. To formulate it in yet another way, the usual requirement of the action being stationary became weaker, since not every variation of the metric $\delta g_{\mu\nu}$ vanishing at infinity can be represented in these terms with also $\delta\phi$ vanishing there \cite{my}.

This idea was later generalised to other conformal and disformal redefinitions of the metric \cite{more}. If all the fluctuations of the physical metric can be smoothly represented in terms of the auxiliary metric which has no extra derivatives in the action, then the variational principle is as strong as it used to be and we have nothing but the General Relativity in disguise. However, if a "gauge" symmetry appears making some parts of the auxiliary metric unphysical, then it entails restrictions for admissible variations of the physical metric because some part of it gets represented only via derivatives of the new fundamental variable $\phi$. And we then obtain the mimetic dark matter again \cite{more}.

An important fact recently noticed in the Paper \cite{new} is that, even in absence of a full-fledged local symmetry for the auxiliary metric, the relation between the two metrics depends on the scalar field $\phi(x)$ and might fail to be bijective at particular configurations of it. Moreover, some configurations like that seem to be unavoidable for almost any transformation with non-trivial dependence on the gradients. It was presented in the Ref. \cite{new} as a blessing, as a benediction to introduce even many more new models of Mimetic Dark Matter and to study the new physics. However, this non-constant rank of the Jacobian matrix of the transformation defining the model seems to me rather as a malediction, with the threats of an ill-defined number of degrees of freedom.

The aim of this paper is to explain the origin of the new mimetic degrees of freedom, hopefully in a pedagogical manner, and to present some worries about viability of these models. In the next section I discuss some fundamental properties of variational principles and illustrate them with simple toy models. In Section 3 then,  I explain what happens with the general disformal transformations in gravity theories. In Section 4 an illustration of that is given in the purely conformal case. Finally, I briefly conclude in the short Section 5.

\section{Making variations in the action}

We work with the Einstein-Hilbert action of General Relativity, $\int d^4 x \sqrt{-g} R(g)$, which is an action depending on only one dynamical variable (with ten components though) $g_{\mu\nu}$. Usually, if the action has derivatives of up to order $n$, we get equations of motion of $2n$-th order. In gravity though, the highest, second-order derivatives enter the action only as a boundary term, and therefore it acts as having the first-order derivatives only and produces the second-order equations of motion.

Very often these higher derivatives in the action are presented as a severe problem for the variational principle. However, if we take it as only a recipe for deriving equations, I do not agree with this opinion. We can perfectly require vanishing of the first variation of the action for arbitrary variations of the metric vanishing at infinity together with one million of its derivatives, or even everywhere outside a big enough radius. Indeed, we cannot demand too many boundary conditions for solving the equations of motion, but nothing prevents us from studying any variations we want, and even variations of finite domain are enough for deriving equations of motion from the stationary action principle.

It's not to say that boundary terms are never of any interest. If we go for a path integral quantisation, for instance, then the canonical number of boundary conditions makes the physical trajectory stationary indeed, in the whole class of paths we integrate over, which can be expected to produce a meaningful quantum theory with a good classical limit then, modulo the common vexatious mathematical troubles. However, this is not the subject of these notes. From now on, I assume that all variations can be taken as smooth and as quickly decaying at infinity as I wish.

\subsection{Toy models}

Let us first start with a simple harmonic oscillator example. Its action can be taken simply as
$$S=\int dt\  ({\dot x}^2-x^2)$$
with the obvious equation of motion: $\ddot x + x=0$. It's a mechanical model with one degree of freedom obeying the second-order equation.

If we make a substitution 
$$x=yz,$$ 
then the action takes the form of
$$S=\int dt\  \left({\dot y}^2 z^2+2{\dot y}y z{\dot z}+{\dot z}^2 y^2-y^2 z^2\right).$$
Performing variations with respect to $y$ and $z$, we get equations which can be easily rewritten as
$$z\cdot \left(\frac{d^2}{dt^2}(yz)+yz\right)=0 \qquad \mathrm{and} \qquad  y\cdot \left(\frac{d^2}{dt^2}(yz)+yz\right)=0$$
respectively.

What we have got is precisely the same harmonic oscillator equation for $x\equiv yz$, with obviously special situations in the $y=0$ and $z=0$ cases. The reason is that the whole action actually depended on $x$ only, and moreover, requiring its variation vanish under arbitrary variations of $y$ and $z$ was absolutely the same condition as vanishing under arbitrary variations of $x$, with no extra troubles coming from the conditions at infinity.

However, suppose we did another redefinition, namely $x=\dot y.$ Then the action gives us a fourth-order equation of motion, or a third-order one for $x\equiv\dot y$ which is more general than the usual second-order harmonic equation, but contains the latter as a specific particular case. More precisely, it is equivalent to putting an arbitrary constant instead of zero to the right hand side of the harmonic equation of motion for the function $x(t)$.

In order to illustrate the reason behind it in a nicer way, let us look at the trivial action of $S=\int dt \cdot x^2$ which obviously has $x=0$ as the equation of motion, with the unique solution of zero $x$, and no freedom in the initial data. However, if I substitute $x=\dot y$, we get $\ddot y=0$, or $\dot x=0$ in terms of the initial variable $x$ as the result. The new set of solutions is an arbitrary constant value of $x(t)$. The old $x=0$ solution is still possible but no longer the only option.

We see that the trick of a substitution with derivatives has widened the class of possible solutions, even for the "physical" variable $x$. Why is that? When we require that $\delta y\to 0$  in the variational procedure for both limits of $t \to \pm\infty$, it means that not every $\delta x$ can be represented in terms of such $\delta y$. Indeed, if $\delta x=\frac{d}{dt}{\delta y}$ with such condition in both the past and the future, it results in $\int dt \cdot \delta x=0$. If we have the variation of the action as $\delta S=\int dt\cdot x \delta x$, it is obvious that an arbitrary constant $x$ would indeed solve for stationarity of the action with respect to the allowed variations of $x$. On the other hand, if for a continuous $x(t)$ there were two moments of time such that $x(t_1)\neq x(t_2)$, then we could find a small enough $\epsilon>0$ and $\delta x(t)$ which is ${\mathcal C}^{\infty}$ smooth, equal to $+1$ in $[t_1-\epsilon\ ,\ t_1+\epsilon]$, equal to $-1$ in $[t_2-\epsilon\ ,\ t_2+\epsilon]$, and equal to $0$ outside both $(t_1-\epsilon-\epsilon^2\ ,\ t_1+\epsilon+\epsilon^2)$ and $(t_2-\epsilon-\epsilon^2\ ,\ t_2+\epsilon+\epsilon^2)$, so that $\delta S\neq 0$.

Of course, working directly in terms of $y(t)$ allows us to get this result of constant $x$ in a much easier way, even if somewhat naive. Having presented this illustration of how it works at the level of variations of $x(t)$, I would like to mention that this is a very important story for physics. For example, if we have an action $\int d^4 x\cdot F_{\mu\nu}F^{\mu\nu}$ for an anti-symmetric tensor which is treated as a fundamental variable, then the equations of motion are simply $F_{\mu\nu}=0$. However, if we substitute $F_{\mu\nu}=\partial_{\mu}A_{\nu}-\partial_{\nu}A_{\mu}$, then it is not only the automatically satisfied homogeneous Maxwell equations but also the condition of $\delta A_{\mu}\to 0$ at infinity what restricts the class of allowed variations of $F_{\mu\nu}$ producing the rest of equations, $\partial_{\mu} F^{\mu\nu}=0$, much more general than the zero tensor.

Coming back to the previous examples, one can also put
$$x={\dot y}z$$
for the harmonic oscillator variable. Derivation of the $z$-equation isn't changed much, and it gives
$${\dot y}\cdot \left(\frac{d^2}{dt^2}({\dot y}z)+{\dot y}z\right)=0$$
producing again the usual harmonic equation for $x(t)$, with a special case of constant $y$, i.e. zero $x$. 

The equation for $y$ goes a bit more general,  
$$\frac{d}{dt} \left(z \cdot \left(\frac{d^2}{dt^2}({\dot y}z)+{\dot y}z\right)\right)=0,$$
and totally in accordance with what we have seen above about substituting $\dot y$ for $x$. However, the equation for $z$ appears restrictive enough to boil the whole story down to the same equation for $x={\dot y}z$ which we had before. This is simply because we can reproduce an absolutely arbitrary variation of $x$ in terms of $y$ and $z$, and it is even enough to vary just $z$ (if not with constant $y$ which anyway means zero $x$) which has no derivatives on it in the definition of $x$.

Note that in their new paper \cite{newer} the Authors of the Ref. \cite{new} also give examples which, in my notations here, can be taken as substitutions of the  $x=z+\dot y$ kind. In particular, this relation does not change anything for the dynamics of $x(t)$ since its arbitrary variation can be given in terms of $\delta z$ which has no extra time derivative. However, as correctly stated in the Ref. \cite{newer}, we can also go for $x=z^3+\dot y$ with another result. In this case, the relation between $x$ and $z$ is still one-to-one (at a fixed choice of $y(t)$). Note though that at the locus of $z=0$, and only there, the linear variations of $x$ are then given only by variations of $\dot y$, what leads to a more general equation for $x(t)$, precisely like it was above for $x=\dot y$. This is a story of an ill-defined number of degrees of freedom in a mechanical system. If we took simply $x=z^3$, the linear variation of the action around $z=0$ would be identically zero.

\subsection{The case of the standard mimetic gravity}

What we see from above is that those variables whose variations can be fully represented in terms of the new variables without derivatives continue with having the very same equations of motion, while those which got substituted by something with derivatives acquire more general equations by virtue of a weaker requirement of the stationary action, i.e. stationary under some restricted class of variations. In a less trivial setup, it can be illustrated in the case of  the standard mimetic gravity \cite{mimetic, my}.

As has been mentioned in the Introduction, the model is defined by
\begin{equation}
\label{standard}
g_{\mu\nu}={\tilde g}_{\mu\nu}{\tilde g}^{\alpha\beta}(\partial_{\alpha}\phi)(\partial_{\beta}\phi).
\end{equation}
Note that, automatically by the definition (\ref{standard}), we have $g^{\alpha\beta}(\partial_{\alpha}\phi)(\partial_{\beta}\phi)=1$. And actually, the model can also be equivalently rewritten \cite{my} as the standard General Relativity with a scalar field having only a Lagrange multiplier term, $\lambda(1-(\partial\phi)^2)$,  in the action.

The Einstein equations then get an effective energy-momentum tensor of the form $\lambda (\partial_{\mu} \phi) (\partial_{\nu} \phi)$. Taking the trace of the new Einstein equations, $G_{\mu\nu}=\lambda (\partial_{\mu} \phi) (\partial_{\nu} \phi)$, and recalling that the scalar field has unit gradient, $(\partial\phi)^2=1$, we see that the value of the Lagrange multiplier is given by the Ricci scalar, $\lambda=-R$. The scalar equation of motion $\bigtriangledown_{\mu}\left(\lambda\partial^{\mu}\phi\right)=0$ takes then the form of $\bigtriangledown_{\mu}\left(R \partial^{\mu}\phi\right)=0$.

To see how it is all related to the classes of variations, let us say it again that the auxiliary metric has a local conformal symmetry under this definition (\ref{standard}). The relation (\ref{standard}) is not invertible. The metrics share the same angles, but an overall factor in the metric $\tilde g$ has no meaning whatsoever, while its counterpart in the physical metric $g$ is represented by the scalar field, or more precisely by its gradients.

If we look at the variation of the Einstein-Hilbert action, $\int d^4x \sqrt{-g}\cdot G^{\mu\nu}\delta g_{\mu\nu}$, the variation of the conformal factor of the metric, $\delta g_{\mu\nu}\propto g_{\mu\nu}$, is responsible for the trace of the field equations. In particular, if we restrain from varying the conformal factor at all, then only the traceless equations of unimodular gravity are obtained. In the current case, a variation of the scalar field gives the conformal type of metric variation as $\delta g_{\mu\nu}={\tilde g}_{\mu\nu}{\tilde g}^{\alpha\beta}\delta\left((\partial_{\alpha}\phi)(\partial_{\beta}\phi)\right)= g_{\mu\nu} g^{\alpha\beta}\delta\left((\partial_{\alpha}\phi)(\partial_{\beta}\phi)\right)$, with the restriction on the class of variations coming from $\delta\phi\to 0$ at infinity. Substituting that into the variation of the action we get $\bigtriangledown_{\alpha}\left(R \partial^{\alpha}\phi\right)=0$ where we have used $G^{\mu}_{\mu}=-R$. That's what we have in mimetic gravity indeed, weaker than $R=0$ of (vacuum) general relativity but stronger than no condition at all in unimodular gravity.

\section{Invertibility of general disformal transformations}

Now let us look at the general disformal transformation. We will express the physical metric $g_{\mu\nu}$ in terms of an auxiliary metric ${\tilde g}_{\mu\nu}$ and a scalar field $\phi$. We denote $X\equiv g^{\mu\nu}(\partial_{\mu}\phi)(\partial_{\nu}\phi)$ and ${\tilde X}\equiv {\tilde g}^{\mu\nu}(\partial_{\mu}\phi)(\partial_{\nu}\phi)$, as well as $\partial^{\mu}\equiv g^{\mu\nu}\partial_{\nu}$ and ${\tilde \partial}^{\mu}\equiv {\tilde g}^{\mu\nu}\partial_{\nu}$. The basic definition then goes as
\begin{equation}
\label{disformal}
g_{\mu\nu}= C(\phi, {\tilde X})\cdot {\tilde g}_{\mu\nu} +  D(\phi, {\tilde X})\cdot (\partial_{\mu}\phi)(\partial_{\nu}\phi)
\end{equation}
with the two arbitrary functions of the two arguments inheriting their names from {\bf C}onformal and {\bf D}isformal parts of the transformation (\ref{disformal}). Note that the standard mimetic gravity case corresponds to $C=\tilde X$ and $D=0$ which entails $X=1$.

As also in the case of mimetic gravity (\ref{standard}), the general shapes of the two metrics are firmly related to each other, and one can solve for the metric $\tilde g$ as
\begin{equation}
\label{inversion}
{\tilde g}_{\mu\nu}= \frac{1}{C} \cdot g_{\mu\nu}-  \frac{ D}{C} \cdot (\partial_{\mu}\phi)(\partial_{\nu}\phi).
\end{equation}
The only problematic point now is to rewrite the arguments of the $C$ and $D$ functions in terms of $X$ instead of $\tilde X$. It was definitely impossible for mimetic gravity since we had the fixed value of $X=1$ while $\tilde X$ was indeed a variable.

Since the disformal part of the transformation (\ref{disformal}) is just a one-dimensional perturbation, one can look for the inverse matrix by making the same perturbation to ${\tilde g}^{\mu\nu}$ and finding the proper coefficient. The result is:
$$g^{\mu\nu}= \frac{1}{C} \cdot\left({\tilde g}^{\mu\nu}-  \frac{ D}{C+D{\tilde X}} \cdot ({\tilde \partial}^{\mu}\phi)({\tilde \partial}^{\nu}\phi)\right).$$
Then we can find the relation between the two gradients squared by simply multiplying the inverse metric by $(\partial_{\mu}\phi)(\partial_{\nu}\phi)$:
\begin{equation}
\label{gradrel}
X=\frac{\tilde X}{C+D{\tilde X}}.
\end{equation}
Obviously, we get $X=1$ indeed for the mimetic case (\ref{standard}) of $C=\tilde X$ and $D=0$.

From the formula (\ref{inversion}) we see that the mapping from one metric to another is bijective if and only if we can invert the $X(\tilde X)$ relation (\ref{gradrel}) to get ${\tilde X}(X)$ instead. Global lack of injectivity with several possible solutions for $\tilde X$ is of no interest for us now, for we only seek this representation  for small fluctuations. Locally the equation (\ref{gradrel}) can be inverted as long as $\frac{dX}{d{\tilde X}}\neq 0$. Strictly speaking, it is a sufficient condition. An isolated point of zero first derivative does not necessarily prevent a mapping from being one-to-one, think of $x^3$ for example. However, such a point will anyway strip some linear fluctuations of $\tilde g$ of their presence at the linear level in the physical metric perturbation $\delta g_{\mu\nu}$, potentially allowing for more freedom in the solutions which correspond to the singular point. And then, in their next paper \cite{newer}, the Authors of the Ref. \cite{new} have used precisely the $x^3$-type trick in order to construct yet another mimetic model with an ill-defined number of degrees of freedom.

Actually, both conditions of non-invertibility (or let's say, singularity conditions, to account for invertible transformations with a zero derivative locus), from the old paper \cite{more} and the one from considering the Jacobian matrix \cite{new}, can be traced back to the very same equation (\ref{gradrel}). Indeed, one can easily find for this function that $\frac{\partial X}{\partial{\tilde X}}=\frac{C+D{\tilde X}-{\tilde X} \frac{\partial}{\partial\tilde X}(C+D{\tilde X})}{(C+D{\tilde X})^2}$ and equate its numerator to zero; or by simply requiring $\frac{\partial}{\partial\tilde X}\left(\frac{1}{X}\right)=0$, we arrive at the condition
$$\frac{\partial}{\partial\tilde X}\left(\frac{C}{\tilde X}+D\right)=0$$
for degeneracy of the Jacobian. This is precisely the condition from both Refs. \cite{more,new}, with the only difference in whether it is taken as a functional equation with the scalar field being fully arbitrary or as a differential equation for particular configurations of the scalar field.

If we want this be true independently of the scalar field configuration, then we obtain the classical relation from the Ref. \cite{more}: 
$$\frac{C}{\tilde X}+D=h(\phi)$$
with an arbitrary function $h$ of the field only, with no dependence on its gradient. The model is then equivalent to the classical mimetic gravity. The value of $X$ does not then depend on $\tilde X$ at all, with possible dependence on $\phi$ though, and the transformation (\ref{disformal}) can never be inverted to find a precise auxiliary metric $\tilde g$ from a given physical metric $g$. 

If, instead of a stably broken case, we are looking for some particular scalar field configurations which would mean only a local singularity of the metric transformation (\ref{disformal}), then $\frac{dX}{d{\tilde X}}=0$ is nothing but the equation from the Refs. \cite{new, newer}
$$C={\tilde X}\frac{\partial}{\partial\tilde X}C+{\tilde X}^2\frac{\partial}{\partial\tilde X}D$$
taken as a differential equation for the scalar field. 

Note that when the relation (\ref{disformal}) is fully invertible (for example $0\neq C=C(\phi)$ and $D=0$) we have nothing but general relativity, with non-invertibility (or singularity) in other cases producing the new degree of freedom of mimetic dark matter. Therefore, having it for some scalar field configurations only must be worrisome if one wants to have a well-defined number of degrees of freedom.

\section{The case of conformal transformations}

For an elementary illustration, let us take the purely conformal case, $D=0$. And for the sake of having it in a more familiar form, as well as automatically arranging for $C\neq 0$, I parametrise it as $C=e^{2f}$. Then we have the relation (\ref{disformal}) as simply
$$g_{\mu\nu}=e^{2f(\phi,\tilde X)}\cdot {\tilde g}_{\mu\nu}.$$
Equations of motion were already derived in the Ref. \cite{new}  relating $\frac{\delta S}{\delta g_{\mu\nu}}$ to $\frac{\delta S}{\delta\tilde g_{\mu\nu}}$ via the Jacobian of the transformation (\ref{disformal}). However, we can also directly substitute the metric $g=g({\tilde g}, \phi)$ into the Einstein-Hilbert action $\int d^4 x \sqrt{-g} R(g)$. Below I will remind the main formulae for that.

The Levi-Civita connection coefficients
$$\Gamma^{\alpha}_{\mu\nu}(g)=\frac12 g^{\alpha\beta}\left(\partial_{\mu}g_{\beta\nu}+\partial_{\nu}g_{\beta\mu}-\partial_{\beta}g_{\mu\nu}\right)$$
would not be changed at all if the function $f$ was constant. Otherwise we get
$$\Gamma^{\alpha}_{\mu\nu}(g)=\Gamma^{\alpha}_{\mu\nu}({\tilde g})+\delta^{\alpha}_{\nu}\partial_{\mu} f +\delta^{\alpha}_{\mu}\partial_{\nu} f - {\tilde g}_{\mu\nu}{\tilde\partial}^{\alpha}f.$$
Considering the standard Riemann tensor
$$R^{\alpha}_{\hphantom{\alpha}\beta\mu\nu}=\partial_{\mu}\Gamma^{\alpha}_{\nu\beta}-\partial_{\nu}\Gamma^{\alpha}_{\mu\beta} +\Gamma^{\alpha}_{\mu\rho}\Gamma^{\rho}_{\nu\beta} -\Gamma^{\alpha}_{\nu\rho}\Gamma^{\rho}_{\mu\beta},$$
defining $\delta\Gamma\equiv \Gamma(g)-\Gamma(\tilde g)$, and taking into account that the Levi-Civita connection is symmetric (and, for later, compatible with its own metric), we get the exact relation
$$R^{\alpha}_{\hphantom{\alpha}\beta\mu\nu}(g)=R^{\alpha}_{\hphantom{\alpha}\beta\mu\nu}(\tilde g)+{\tilde \bigtriangledown}_{\mu}\delta\Gamma^{\alpha}_{\nu\beta}-{\tilde \bigtriangledown}_{\nu}\delta\Gamma^{\alpha}_{\mu\beta} +\delta\Gamma^{\alpha}_{\mu\rho}\cdot \delta\Gamma^{\rho}_{\nu\beta} -\delta\Gamma^{\alpha}_{\nu\rho}\cdot\delta\Gamma^{\rho}_{\mu\beta}$$
which is all we need.

After some rather elementary calculations, we obtain the Ricci tensor $R_{\mu\nu}\equiv R^{\alpha}_{\hphantom{\alpha}\mu\alpha\nu}$
$$R_{\mu\nu}(g)=R_{\mu\nu}({\tilde g})-2{\tilde \bigtriangledown}_{\mu}\partial_{\nu}f+2(\partial_{\mu}f)(\partial_{\nu}f) -\left({\tilde\square} f+2({\tilde\partial}f)^2\right){\tilde g}_{\mu\nu},$$
the curvature scalar $R\equiv g^{\mu\nu}R_{\mu\nu}$
\begin{equation}
\label{scaltr}
R(g)=e^{-2f}\cdot\left(R({\tilde g})-6{\tilde\square} f-6({\tilde\partial}f)^2\right),
\end{equation}
and the Einstein tensor $G_{\mu\nu}\equiv R_{\mu\nu}-\frac12 R g_{\mu\nu}$
\begin{equation}
\label{stonetr}
G_{\mu\nu}(g)=G_{\mu\nu}({\tilde g})-2{\tilde \bigtriangledown}_{\mu}\partial_{\nu}f+2(\partial_{\mu}f)(\partial_{\nu}f) +\left(2{\tilde\square} f+({\tilde\partial}f)^2\right){\tilde g}_{\mu\nu}
\end{equation}
relations.

Substituting the curvature scalar (\ref{scaltr}), and $ \sqrt{- g}=e^{4f} \sqrt{-\tilde g}$, into the Einstein-Hilbert action and integrating by parts in the $e^{2f}{\tilde\square} f$ term, we finally get
\begin{equation}
\label{action}
S=\int d^4x \sqrt{-g}\cdot R(g)=\int d^4x \sqrt{-\tilde g}\cdot e^{2f}\left(R({\tilde g})+6 ({\tilde\partial}f)^2 \right).
\end{equation}
Let's prepare a couple of convenient formulae concerning the variations of the action (\ref{action}). 

First, I denote by $\delta^{\prime}S$ the partial variation of the action (\ref{action}) with respect to the metric ${\tilde g}$, neglecting the variation of the argument of $f$. The result would be the full metric variation $\delta_{\tilde g} S$ if the function $f$ depended only on something else:
\begin{multline}
\label{partvar}
\delta^{\prime}S=\int d^4 x \sqrt{-\tilde g}\cdot \delta {\tilde g}^{\mu\nu}\cdot \left(G_{\mu\nu}({\tilde g})+6(\partial_{\mu}f)(\partial_{\nu}f)-3({\tilde\partial}f)^2{\tilde g}_{\mu\nu}+\left({\tilde g}_{\mu\nu}{\tilde\square}-{\tilde \bigtriangledown}_{\mu}{\tilde \bigtriangledown}_{\nu}\right)\right)e^{2f}\\
=\int d^4 x \sqrt{-\tilde g}\cdot G_{\mu\nu}(g)e^{2f} \cdot \delta{\tilde  g}^{\mu\nu}
\end{multline}
where for the first equality we used that ${\tilde g}^{\mu\nu}\delta R_{\mu\nu}({\tilde g})={\tilde g}_{\mu\nu}{\tilde\square}\delta{\tilde g}^{\mu\nu}-{\tilde \bigtriangledown}_{\mu}{\tilde \bigtriangledown}_{\nu}\delta{\tilde g}^{\mu\nu}$, and for the second equality we explicitly acted with the differential operator we had there on $e^{2f}$ and used the relation (\ref{stonetr}) between the Einstein tensors.

Second, the variation of the action (\ref{action}) with respect to the function $f$, pretending that it was a fundamental variable in itself, gives
\begin{multline}
\label{funvar}
\delta_{f}S=\int d^4 x \sqrt{-\tilde g}\cdot\left(2R({\tilde g})-12({\tilde\partial}f)^2-12{\tilde\square}f\right)e^{2f}\cdot\delta f\\
= 2\int d^4 x \sqrt{-\tilde g}\cdot R({ g})e^{4f}\cdot\delta f=2\int d^4 x \sqrt{- g}\cdot R({ g})\cdot\delta f
\end{multline}
with the standard relation (\ref{scaltr}) having been used.

\subsection{The $\tilde X$-free case}

In the case of 
$$C=C(\phi)=e^{2f(\phi)},$$ 
without the $\tilde X$ argument, there is absolutely no dependence on the metric inside the function $f$. And this is a purely invertible case, with the metrics (and their linear variations) being in a one-to-one correspondence to each other (assuming some particular, but arbitrary, scalar field $\phi(x)$ configuration).

In this case
$$\delta_{\tilde g}S=\delta^{\prime}S=\int d^4 x \sqrt{-\tilde g}\cdot G_{\mu\nu}(g)e^{2f} \cdot \delta{\tilde  g}^{\mu\nu}$$
with the standard Einstein equations $G_{\mu\nu}(g)=0$ reproduced. And this is of course all the information in the system, with the field $\phi$ having no more meaning than taking part in constructing the physical variable $g$ out of the fundamental variables $\tilde g$ and $\phi$.

The variation of $\phi$ gives $\delta f= f^{\prime}\delta \phi$. And as long as $f^{\prime}\neq 0$, the variation $\delta_{\phi}S$ of the action (\ref{action}) yields the equation $R(g)=0$ which is precisely the trace of the Einstein equations. This is nothing new but rather natural since $\delta\phi$ can also describe the conformal variations of the physical metric $g_{\mu\nu}$, and as it does so without any extra derivatives, the equation has not become a weaker one. 

In other words, in the metric-free case of $f=f(\phi)$, the action (\ref{action}) simply reproduces the general relativity, in quite an intricate shape though.

\subsection{The general case}

If we now consider the general case, then the variation of the auxiliary metric also produces 
$$\delta_{\tilde g} f= f_{\tilde X} \delta_{\tilde g}{\tilde X} =f_{\tilde X}(\partial_{\mu}\phi)(\partial_{\nu}\phi) \delta{\tilde g}^{\mu\nu}$$ 
with $ f_{\tilde X}\equiv\frac{\partial f}{\partial\tilde X}$. We then get from equations (\ref{partvar}) and (\ref{funvar}) the variation
$$\delta_{\tilde g}S=\delta^{\prime}S+\frac{\delta S}{\delta f}\delta_{\tilde g} f=\int d^4 x \sqrt{-\tilde g}\cdot\left( G_{\mu\nu}(g)e^{2f}+2R(g)e^{4f} f_{\tilde X}(\partial_{\mu}\phi)(\partial_{\nu}\phi)\right)\cdot \delta{\tilde  g}^{\mu\nu}$$
and the equation of motion
$$G_{\mu\nu}(g)+2e^{2f}f_{\tilde X}R(g)\cdot (\partial_{\mu}\phi)(\partial_{\nu}\phi)=0$$
reproducing the classical case of $G_{\mu\nu}(g)+R(g)\cdot (\partial_{\mu}\phi)(\partial_{\nu}\phi)=0$ for the mimetic dark matter \cite{mimetic,my} when $f=\frac12 \log \tilde X$. Of course, the equation looks more general than that of Einstein, but it has all the general relativistic cases as solutions with $G_{\mu\nu}(g)=0$ and therefore $R(g)=0$, or it can also come as a constant field $\phi$ case if $X=0$ is allowed.

Taking the trace of this equation with $g^{\mu\nu}=e^{-2f}\cdot {\tilde g}^{\mu\nu}$, we get 
$$\left(-1+2 f_{\tilde X}\tilde X\right)\cdot R(g)=0.$$ 
Then either the equation does not actually have any new solution due to the need of $R(g)=0$, and the variations of $\tilde g_{\mu\nu}$ have reproduced all the possible variations of $g_{\mu\nu}$, or we have $ f_{\tilde X}=\frac{1}{2\tilde X}$. Since $C=e^{2f}$, the latter means that $ C_{\tilde X}=\frac{C}{\tilde X}$ which is precisely the condition of $\frac{dX}{d\tilde X}=0$ for the function $X(\tilde X)$ defined by the equation (\ref{gradrel}) with $D=0$. We can either solve it as $C(\phi,\tilde X)= h(\phi)\cdot \tilde X$ and have the usual mimetic gravity of the Ref. \cite{more}, or we can go for a more general case, and then it gives us only general relativistic solutions, except for the singular values of $\tilde X$ with  $ f_{\tilde X}=\frac{1}{2\tilde X}$, or $C=C_{\tilde X}\tilde X$ like in the Ref. \cite{new}, which then allow for the dark-matter-like additions. In particular, for the case of $C=e^{\tilde X -1}$ from that paper, we confirm the requirement of $\tilde X=1$.

Note that we can also easily find the equation of $\phi$ from the variation (\ref{funvar}) because
$$\delta_{\phi} f=f_{\phi}\delta\phi+2f_{\tilde X}{\tilde g}^{\mu\nu}(\partial_{\mu}\phi)(\partial_{\nu}\delta\phi)=f_{\phi}\delta\phi+2e^{2f}f_{\tilde X}{g}^{\mu\nu}(\partial_{\mu}\phi)(\partial_{\nu}\delta\phi).$$
In particular, for $f=f(\tilde X)$ we simply have 
$$\bigtriangledown_{\nu}\left(f^{\prime} e^{2f} R(g) \partial^{\nu}\phi\right)=0$$
which obviously reproduces the standard mimetic case of  $\bigtriangledown_{\mu}\left(R(g) \partial^{\mu}\phi\right)=0$ for $f=\frac12 \log \tilde X$.

\subsection{Solvability in terms of $\tilde g$ and $\phi$}

We have presented the equations of motion in terms of the physical metric $g_{\mu\nu}$ and the scalar field $\phi$. In case of the transformation (\ref{disformal}) being not invertible, it is not yet the whole story. If we find a solution of these equations, we will also have to check whether the result obtained in this way can be written in terms of ${\tilde g}_{\mu\nu}$. In particular, in the classical mimetic ($C=\tilde X$ and $D=0$) case of Refs. \cite{mimetic,my}, the solution must have $X=1$, for otherwise no auxiliary metric would be able to produce such a configuration, according to the relation (\ref{gradrel}).

In general cases, there is no automatic condition of the $X=1$ type. However, the loci of $\frac{dX}{d\tilde X}=0$ might either give one-sided restrictions (inequalities) on possible values of $X$; or when it is an even order root of the $X_{\tilde X}$ derivative, a proper auxiliary metric might still exist for the physical one with any values of $X$, but possibly with singularities in its derivatives via a singular $\frac{d\tilde X}{dX}$ value, see also the Ref, \cite{newer}. 

At the same time, all the equations of motion have the factor of $R(g)$ in front of the scalar quantities. It means that if we take any Einstein spacetime with $G_{\mu\nu}(g)=0$, then modulo the solvability for $\tilde X$, practically any configuration of $\phi(x)$ might be a solution. Moreover, the linearised equations around such a background do not then feature the variation of the scalar $\delta\phi$. It can be taken as a version of strong coupling which I would call "extreme freedom".

Having the non-constant rank of the Jacobian of the transformation (\ref{disformal}) also adds to these worries at higher orders in perturbations. With an Einstein spacetime ($G_{\mu\nu}(g)=0$) background, one can take many different $\phi(x)$. If it was far away from $C=C_{\tilde X}\tilde X$, then even at higher orders of perturbations we would have only pure gravity solutions, while otherwise the mimetic dark matter options would probably pop out, too. All in all, these models do call for better investigation of their fundamental properties.

\subsection{The role of the derivatives}

Finally, let me stress it once more that the crucial aspect of the changes of variables which have produced more general equations of motion is the presence of new derivatives being substituted into the action. Instead of the scalar gradient $\partial_{\mu}\phi$ we could have used a vector $A_{\mu}$, with the 
$${\tilde Y}={\tilde g}^{\mu\nu}A_{\mu}A_{\nu}$$ 
argument in the functions $C$ and $D$ of the transformation (\ref{disformal}), in place of $\phi$ and $\tilde X$. Possessing no new derivatives, it wouldn't lead us beyond General Relativity. 

Indeed, let's take the conformal case with $f=f(\tilde Y)$. For the type of dependence on the metric, there is no difference between $\tilde X$ and $\tilde Y$, and we get the same $\tilde g$-equation as before, with $A_{\mu}$ instead of $\partial_{\mu}\phi$:
$$G_{\mu\nu}(g)+2e^{2f}f_{\tilde Y}R(g)\cdot A_{\mu}A_{\nu}=0.$$
This equation is again more general than the simple $G_{\mu\nu}(g)=0$ when $f_{\tilde Y}= \frac{1}{2\tilde Y}$, due to precisely the same reason as before: the variation $\delta {\tilde g}_{\mu\nu}$ cannot reproduce all the possible variations $\delta g_{\mu\nu}$ of the physical metric if there is a value of $\tilde Y$ such that $\frac{dY}{d\tilde Y}=0$.

However, now the vector field takes the role of producing the missing variations, and it has no extra derivatives on it. Then we see that
$$\delta_A f= 2f_{\tilde Y} {\tilde g^{\mu\nu}}A_{\mu}\delta A_{\nu}$$
which together with the variation formula (\ref{funvar}) gives us the simple
$$R(g)A_{\mu}=0$$
equation. It is either $R(g)=0$, or $A_{\mu}=0$ which anyway leads then to $G_{\mu\nu}(g)=0$ and therefore $R(g)=0$. In other words, the scalar field model had had novel solutions only due to the presence of derivatives in the change of variables (\ref{disformal}).

\section{Conclusions}

I have shown why the disformal transformations very often lead to new solutions of mimetic gravity type. I fully confirm the main findings of the recent paper \cite{new}, and I think that I have managed to explain the workings of these models in a somewhat more elementary way. At the same time, the foundational properties of such theories look very worrisome and might often be incompatible with even purely mathematical well-posedness. This issue definitely deserves further investigation.

\end{document}